\documentclass[english]{article}
\usepackage[T1]{fontenc}
\usepackage[latin1]{inputenc}
\usepackage{graphicx}
\usepackage{setspace}
\usepackage{amssymb}

\makeatletter

\providecommand{\LyX}{L\kern-.1667em\lower.25em\hbox{Y}\kern-.125emX\@}
\newcommand{\noun}[1]{\textsc{#1}}

 \newcommand{\lyxaddress}[1]{
   \par {\raggedright #1 
   \vspace{1.4em}
   \noindent\par}
 }

\usepackage{babel}
\makeatother
\begin{document}

\title{Causality Violation in Non-local Quantum Field Theory}

\author{Ambar Jain and Satish D. Joglekar}

\maketitle

\lyxaddress{Department of Physics, I.I.T.Kanpur, Kanpur 208016 (INDIA)}

\begin{abstract}
We study the causality violation in the non-local $\phi ^{4}$-theory
(as formulated by Kleppe and Woodard) containing a finite mass scale
$\Lambda $. Starting from the Bogoliubov-Shirkov criterion for causality,
we construct and study combinations of S-matrix elements that signal violation of causality in the one loop approximation. We find that
the causality violation in the exclusive process $\phi +\phi \rightarrow \phi +\phi $
grows with energy, but the growth with energy, (for low to moderate
energies) is suppressed to all orders compared to what one would expect
purely from dimensional considerations. We however find that the causality
violation in other processes such as $\phi +\phi \rightarrow \phi +\phi +\phi +\phi $
grows with energy as expected from dimensional considerations at low
to moderate energies. For high enough energies comparable to the mass
scale $\Lambda $, however, we find a rapid (exponential-like) growth
in the degree of causality violation. We suggest a scenario, based
on an earlier work, that will enable one to evade a large theoretical
causality violation at high energies, should it be unobserved experimentally.
\end{abstract}

\section{\noun{introduction} }

The chief problem one is faced with any local quantum field theory
(LQFT) is the presence of ultra-violet divergences. A number of regularization
procedures have been proposed to deal with such divergences. The basic
problem of divergences arises from the ill-defined nature of a product
of two local field operators at the same space-time point \cite{iz}
in a LQFT. One of the natural suggestions to deal with this problem
therefore has been that we should attempt to replace a local interaction
by a nonlocal (smeared) interaction which somehow would take care of the divergences
arising from two local field operators at the same space-time point
\cite{book}. While making such an attempt, however, there is a danger
that one may loose physically meaningful properties of the local theory
such as unitarity, stability \cite{ew,ew2}, gauge invariance (in
the case of gauge theories) and causality \cite{kw}. Many attempts
have been made in this direction to construct a finite and physically
meaningful quantum field theory with the use of nonlocal interactions
\cite{book,kw}. One of the physical requirements of the physical theory
is the unitarity of the S-Matrix. This requires that in the process
of introducing nonlocality, one does not introduce any unphysical
ghost degrees of freedom that can contribute to an S-matrix.  Also, for local
theories with a local gauge-invariance, it is necessary to introduce
nonlocal interactions in a manner that is compatible with the local
gauge symmetry (or an equivalent non-local symmetry), so that the
essential consequences of a local gauge symmetry are preserved after
non-localization.

An elegant formulation of nonlocal quantum field theory that preserves
unitarity and (an equivalent of) local gauge symmetry has been
given by Kleppe and Woodard \cite{kw}. It was based upon the earlier works by Moffat\cite{m90}, Elizer
and Woodard \cite{ew,ew2} and on an explicit construction for QED by Evens et al \cite{e91}. A
systematic procedure for constructing a non-local action from a local
QFT has been presented in Ref. \cite{kw}. Equally importantly, this work has established that the local/global symmetries can be
preserved in their nonlocal form and the WT identities of a local
QFT derivable from local symmetries such as gauge invariance/BRS symmetry
find their natural nonlocal extensions (at least in the Feynman gauge).
It has since been studied extensively \cite{kw2,ext}. It has also been
found useful for regularizing theories with supersymmetry \cite{susy}
(where dimensional regularization does not work) and for the regularization
of BV theories \cite{pt}.

This formulation of a non-local field theory can be looked upon either
as a regularization \cite{e91,kw2} or as a physical theory with a
finite mass parameter $\Lambda $ \cite{m90,kw}. In either case,
and for \emph{any finite $\Lambda $}, the theory is unitary and has
an appropriate generalization of a local gauge symmetry. The latter
of these interpretations can further be looked upon in a number of
distinct ways. One could think the non-locality as representing a
form factor with a momentum cut-off $\Lambda $\cite{m90}. One could
also think of this theory as embodying a granularity of space-time
of the scale $1/\Lambda $ or as an intrinsic mass scale $\Lambda $
\cite{kw,js,j01,bj}. One could also consider such a theory as representing
an effective field theory valid when the energy scale involved is smaller than $\Lambda $
\cite{j01_2}.

In an earlier work \cite{j01}, we have found such a non-local formulation
\emph{with a finite $\Lambda $,} very useful in understanding the
renormalization program in the renormalizable field theories. We have
shown that this formulation enables one to construct a mathematically
consistent framework in which the renormalization program can be understood
in a natural manner. The framework does not require any violations
of mathematical rigor usually associated with the renormalization
program. This framework, moreover, made it possible to theoretically
estimate the mass scale $\Lambda $. The nonlocal formulations can
also be understood \cite{j01_2} as an effective field theory formulation of a physical
theory that is valid up to mass scale $\sim $$\Lambda $. In such
a case, the unknown physics at energy scales higher than $\Lambda $ {[}such
as a structure in terms of finer constituents, additional particles,
forces, supersymmetry etc {]} can \emph{effectively be represented} in a \emph{consistent}
way (a unitary gauge-invariant finite (or renormalizable)  theory)
by the non-local theory. In other words, the nonlocal standard model can serve as such an
effective field theory \cite{j01_2} and will afford a model-independent
way of reparametrizing the effects beyond standard model consistently. 

Despite the various advantages of these nonlocal formulations, it
suffers from some limitations: such a theory, while it preserves causality
at the tree-level S-matrix (which is the \emph{same as} the local one), has
quantum violations of causality, expected to be small by a factor
$\sim $$\frac{g^{2}}{16\pi ^{2}}$\label{cornish/kw}. In this work,
we shall carry out an elementary study of the question of causality
violation (CV) in NLQFT . We shall adopt the approach of Bogoliubov
and Shirkov \cite{bs}, and use Bogoliubov-Shirkov criterion \cite{bs1}
to study CV. Bogoliubov-Shirkov criterion (See section 2.3) allows
one obtain certain combinations of S-matrix elements that must vanish
if causality is preserved; and therefore these combinations constitute
quantities that can characterize the CV in a NLQFT. We study these
in the 1-loop approximation for the $\lambda $$\phi ^{4}$ theory,
this being the simplest field theory model where definite conclusions
can be drawn.

We shall now introduce the plan of the paper. In section 2, we shall introduce the formalism of non-local field theories for the $\lambda \phi^4$ theory and give Feynman rules. We shall also summarize the causality condition as formulated by Bogoliubov and Shirkov. We shall construct quantities that can be used to characterize the causality violation. We shall also make brief comments on the possible violation of causality in LQFT \cite{heg95} and its distinction from the causality violation we shall address to in the present context of the NLQFT. We shall also summarize the scenario suggested in \cite{j01_2}. In Section 3, we shall evaluate the causality violation effects at the one loop level in two of the exclusive processes: (i) 2 scalars $\rightarrow$  2 scalars, (ii) 2 scalars $\rightarrow$   4 scalars. We shall show that at an energy scale small compared to $\Lambda $, the causality violation in the first process is of an order smaller than what would be expected from purely dimensional consideration. We shall further show that at energies comparable to $\Lambda $ or greater, we shall find an exponential-like growth in the causality violation. For the second process, we find that we have a causality violation as expected from general dimensional arguments for energy scales $<<\Lambda$. On the other hand, at energy scales $\sim \Lambda $ we again find an exponential-like growth. In section 4, we shall give a general argument to show that the results noted in the one loop for the  2 scalars $\rightarrow$  2 scalars aught to be valid to all orders.  In section 5, we shall present a scenario based on that presented in \cite{j01_2}, that allows one to evade a large scale causality violation should it be unobserved experimentally.

\section{{\normalsize PRELIMINARIES}}

In this section, we shall summarize, for our future use, various known
results on the non-local field theories, Bogoliubov-Shirkov criterion
of causality, and brief comments on possible causality violation in
LQFT and the pertinence (or its lack) of these results to the present
discussion. We will also summarize the view-point regarding the non-local field theories as presented in \cite{j01_2}.

\subsection{{\normalsize NON-LOCAL REGULARIZATION: }}

We shall first review, very briefly, the construction of a non-local
field theory with a finite mass scale $\Lambda $, given its local counterpart. We shall present
the construction with reference to the $\lambda \phi ^{4}$ theory as
presented in \cite{kw}; since that suffices for our purpose. 

We start with the local action for a field theory, in terms of a generic
field $\phi $, as the sum of the quadratic and the interaction part:
\[
S[\phi ]=F[\phi ]+I[\phi ]\]
and express the quadratic piece as \[
F[\phi ]=\int d^{4}x\phi _{i}(x)\Im _{ij}\phi _{j}(x)\]
We define the regularized action in terms of the smeared field $\widehat{\phi }$,
defined in terms of the kinetic energy operator $\Im _{ij}$ as,\[
\widehat{\phi }\equiv \mathcal{E}^{-1}\phi \, \, \, \, \, \, \, \, \, \, \, \mathcal{E}\equiv \exp [\Im /\Lambda ^{2}]\]
The nonlocally regularized action is constructed by first introducing
an auxiliary action $S[\phi ,\psi ]$. It is given by \[
S[\phi ,\psi ]=F[\hat{\phi }]-A[\psi ]+I[\phi +\psi ]\]
where $\psi $ is called a {}``shadow field'' with an action\[
A[\psi ]=\int d^{4}x\psi _{i}O_{ij}^{-1}\psi _{j}\]
 with $O$ defined by \[
O\equiv \frac{\mathcal{E}^{2}-1}{\Im }\]
The action of the non-local theory is defined as\[
\hat{S}[\phi ]=S[\phi ,\psi ]\Vert _{_{\psi =\psi [\phi ]}}\]
 where $\psi [\phi ]$ is the solution of the classical equation\[
\frac{\delta S}{\delta \psi }=0\]
The Feynman rules for the nonlocal theory are simple extensions of the local
ones. The vertices are unchanged but every leg can connect either
to a smeared propagator\[
\frac{i\mathcal{E}^{2}}{\Im +i\epsilon }=-i\int _{1}^{\infty }\frac{d\tau }{\Lambda ^{2}}exp\{\frac{\Im \tau}{ \Lambda ^{2}}\}\]
or to a shadow propagator {[}shown by a line crossed by a bar{]}

$$\frac{i[1-\mathcal{E}^{2}]}{\Im +i\epsilon }=-iO =-i \int^{1}_0\frac{d\tau }{\Lambda ^{2}}exp\{\frac{\Im \tau}{ \Lambda ^{2}}\}$$

In the context of the $\lambda \phi^4$ theory, we have,
$$\Im = -\partial^2-m^2 \quad I(\phi)= -\frac{\lambda}{4}  \phi^4$$

The Feynman rules for propagator in momentum space read:
\begin{enumerate}   

\item For the $\phi $-propagator (smeared propagator) denoted by a straight line: $i\frac{\left\{ exp\left[\frac{p^{2}-m^{2}}{\Lambda ^{2}}\right]\right\} }{p^{2}-m^{2}+ie}$

\item For the $\psi $-propagator denoted by a \emph{ barred} line:$$i\frac{\left\{ 1-exp\left[\frac{p^{2}-m^{2}}{\Lambda ^{2}}\right]\right\} }{p^{2}-m^{2}+ie}$$

\item The 4-point vertex is as in the local theory, except that any of the lines emerging from it can be of either type. (There is accordingly a statistical factor).

\item In a Feynman diagram, the internal
lines can be either shadow or smeared, with the exception that no diagrams can have closed shadow loops. \end{enumerate}

\subsection{{\normalsize Condition of Causality}}

The causality condition that we have used to investigate causality
violation in NLQFT is the one discussed by Bogoliubov and Shirkov \cite{bs}.
They have shown that an S-matrix for a theory that preserves causality
must satisfy the condition\begin{equation}
\begin{array}{ccc}
 \frac{\delta }{\delta g(x)}\left(\frac{\delta S[g]}{\delta g(y)}S^{\dagger }[g]\right)=0 & for & x<\sim y\end{array}
\label{eq:CC}\end{equation}
 where $x<\sim y$ means that either $x^{0}<y^{0}$ or $x$ and $y$
are space like separated. The above condition has been formulated
treating the coupling $g(x)$ as space-time dependent. This is used together with the unitarity
condition \begin{equation}
S^{\dagger }[g]S[g]=1\label{eq:unit}\end{equation}
 to obtain the form of the S-matrix in LQFT. We shall use them together to study the
causality violation in NLQFT. To extract the physics from the above
mentioned causality condition, in the perturbative sense, S-matrix
is written as a functional Taylor expansion in powers of coupling
$g(x)$: \begin{equation}
S[g]=1+\sum _{n\geq 1}\frac{1}{n!}\int S_{n}(x_{1},...,x_{n})g(x_{1})...g(x_{n})dx_{1}...dx_{n},\label{eq:CCexp}\end{equation}
 in which $S_{n}(x_{1},...,x_{n})$ are \emph{operator} expressions (symmetric
in all arguments) which depend upon the field operators and on their
partial derivatives at the points $x_{1},...,x_{n}.$ We note that
experimental information is not however about an $S_{n}(x_{1},...,x_{n})$
but about\[
S_{n}\equiv \int S_{n}(x_{1},...,x_{n})dx_{1}...dx_{n}\] obtained by putting $g(x)=constant=g$. As shown in \cite{bs1}, one can obtain
perturbatively useful causality conditions for
each $n=1,2,3,...$ from (\ref{eq:CC}) and (\ref{eq:unit}) and using the expansion
(\ref{eq:CCexp}): 
\begin{eqnarray}
H_{n}(y,x_{1},...,x_{n})=iS_{n+1}(y,x_{1},...,x_{n}) &  & \nonumber \\
+i\sum _{0\leq k\leq n-1}P\left(\frac{x_{1},...,x_{k}}{x_{k+1},...,x_{n}}\right)S_{k+1}(y,x_{1},...,x_{k})S_{n-k}^{\dagger }(x_{k+1},...,x_{n}) & = & 0\label{eq:crit}
\end{eqnarray}
 Here, $P\left(\frac{x_{1},...,x_{k}}{x_{k+1},...,x_{n}}\right)$
is defined as the sum over the distinct ways of partitioning ($\frac{n!}{k!(n-k)!} $ in number)  $\{x_{1},x_{2},x_{3},........x_{n}\}$
into two sets of $k$ and $(n-k)$ (such as $\{x_{1},x_{2},x_{3},........x_{k}\}$ $\{x_{k+1},........x_{n}\}$).
Similarly perturbatively useful unitarity condition, for each n,
is given by\begin{eqnarray*}
S_{n}(x_{1},...,x_{n})+S_{n}^{\dagger }(x_{1},...,x_{n}) &  & \\
+\sum _{1\leq k\leq {n-1}}P\left(\frac{x_{1},...,x_{k}}{x_{k+1},...,x_{n}}\right)S_{k}(x_{1},...,x_{k})S_{n-k}^{\dagger }(x_{k+1},...,x_{n}) & = & 0.
\end{eqnarray*}
 For $n=1,2$ respectively, causality condition is explicitly given
by\begin{equation}
H_{1}(x,y)\equiv iS_{2}(x,y)+iS_{1}(x)S_{1}^{\dagger }(y)=0\label{causal1}\end{equation}
\begin{equation}
H_{2}(x,y,z)\equiv iS_{3}(x,y,z)+iS_{1}(x)S_{2}^{\dagger }(y,z)+iS_{2}(x,y)S_{1}^{\dagger }(z)+iS_{2}(x,z)S_{1}^{\dagger }(y)=0\label{causal2}\end{equation}
 and unitarity condition is given by\begin{equation}
S_{1}(x)+S_{1}^{\dagger }(x)=0\label{unitary1}\end{equation}
 \begin{equation}
S_{2}(x,y)+S_{2}^{\dagger }(x,y)+S_{1}(x)S_{1}^{\dagger }(y)+S_{1}(y)S_{1}^{\dagger }(x)=0\label{unitary2}\end{equation}
 The unitarity conditions could be used with causality condition to
replace $S_{n}^{\dagger }$ factor in (\ref{eq:crit}) by appropriate
$S_{m}$ terms. One can then evaluate the matrix elements of $H_{n}(y,x_{1},...,x_{n})$
between appropriate initial and the final states to convert the relations
(\ref{eq:crit}) in terms of the S-matrix \emph{element coefficients} in (\ref{eq:CCexp}). In the
case of the local theory, these relations are trivially satisfied.
In the case of the nonlocal theories, these quantities, on the other
hand, afford a way of characterizing the causality violation. Further,
these quantities contain however not the usual S-matrix elements that one
can observe in an experiment (which are obtained with a \emph{constant
i.e. space-time-independent} coupling), but rather the coefficients
in (\ref{eq:CCexp}). We thus find it profitable to construct appropriate
space-tine integrated versions out of $H_{n}(y,x_{1},...,x_{n})$.
Thus, for example, we can consider
\begin{equation}    
H_{1}\equiv \int d^{4}x\int d^{4}y[\vartheta (x_{0}-y_{0})H_{1}(x,y)+\vartheta (y_{0}-x_{0})H_{1}(y,x)]\label{eq:H_1} \end{equation}
The causality condition (\ref{causal1}) would then imply that a lack
of CV requires that\footnote{These statements are subject to the counterterms in the definition of $H_1$ to be discussed shortly.}  $H_{1}\equiv 0$. Conversely, a non-zero $H_{1}$
necessarily implies CV. Recalling that $S_{2}(x,y)=S_{2}(y,x)$, we
can write\[
H_{1}=i\int d^{4}x\int d^{4}y S_{2}(x,y)-i\int d^{4}x\int d^{4}yT[S_{1}(x)S_{1}(y)]\]
 which can be expressed entirely in terms of Feynman diagrams that
appear in the usual S-matrix amplitudes. In a similar manner, we can
formulate
\begin{eqnarray}  
H_{2} & \equiv &  \int d^{4}x\int d^{4}y\int d^{4}zH_{2}(x,y,z)\vartheta (x_{0}-y_{0})\vartheta (y_{0}-z_{0}) \\
&  &      +5\,\mbox{symmetric}\; \mbox{terms}\label{eq:H_2} 
\end{eqnarray} 
and can itself be expressed in terms of Feynman diagrams. 

Finally, we comment on the counterterms in the above equations. When we wrote the original causality condition in terms of $H_1(x,y)$, with $y<\sim x$, $x$ and $y$
 were distinct points and there were no divergences in both the terms that are present in $H_1(x,y)$. On the other hand, when we convert it to $H_1$ defined above, (each term in its definition now allows for coincident points), and contains a "subtraction" or a counterterm. The counterterm\footnote{The expansion of (\ref{eq:CCexp}) is the expansion of the \emph{renormalized} S-matrix in terms of the \emph{renormalized} coupling $g$.}  \emph{present} in $\int d^{4}x\int d^{4}y S_{2}(x,y)$ makes this term finite while the "subtraction" that \emph{has} to be included in $\int d^{4}x\int d^{4}y T[S_{1}(x)S_{1}(y)]$ has to be such that as $\Lambda \rightarrow \infty$ (i.e. the local limit) $H_1$ must vanish. This criterion, however, does not uniquely determine the net counterterm in $H_1$; there is an ambiguity of a \emph{constant, i.e. momentum-independent} term that vanishes as $\Lambda \rightarrow \infty $. We shall assume, for definiteness, that experiments show no causality violation at low energies, and we shall adjust this counterterm accordingly. Nonetheless, the presence of a \emph{momentum-dependent} CV terms in $H_1$ necessarily signals CV at \emph{some} energy scale, irrespective of the ambiguity in the counterterm.

\subsection{{\normalsize Causality Violation in Local Quantum Field Theories\label{cvlqft}}}

We shall briefly discuss the issue of causality violation in the LQFT
and point out the distinction between the issue there and here in
the context of the NLQFT. There is much discussion in literature about
the CV in LQFT and is thought to be still a controversial matter \cite{heg95} .
However, these discussions revolve around processes that take place
in a \emph{finite duration,} unlike an S-matrix which relates to the
time-evolution of a scattering system over an infinite duration. The
possible signal of CV in LQFT in fact vanishes as the time of observation
of a process goes to infinity. This is in fact consistent with no
CV in LQFT according to the BS criterion; which refers to an S-matrix
element.

In light of Bogoliubov and Shirkov formulation of S-matrix theory
\cite{bs1} and construction of $S_{n}$, which is widely used in
perturbative QFT, it is clear that causality violation does not occur
in the S-matrix theory in a LQFT. Therefore we have studied CV in NLQFT
in the S-matrix formalism.

\subsection{Non-local Field Theory as an Effective Field Theory}

In this subsection, we shall summarize the view-point presented in
references \cite{j01_2} and \cite{j01} regarding the non-local theories
as effective field theories valid up to a mass scale $\Lambda $. The
basic idea presented in \cite{j01_2} was with reference to a quantum
field theory for which the LSZ formulation leads to the condition
$0<Z<1$. In a local quantum field theory, this condition is ignored
in perturbation theory as the wave-function renormalization $Z$ diverges.
It was suggested in \cite{j01_2} that we can resurrect this condition
in a NLQFT with an intrinsic scale $\Lambda $, and in fact give a
meaning to the scale and make an estimate using the condition $0<Z<1$
\emph{as applied in perturbation theory.} The idea was to require
that the mass-scale $\Lambda $ (together with the coupling) be such
that the calculated $Z$ satisfies $0<Z<1$. This lead to a condition
of the type \[
\frac{g^{2}}{16\pi ^{2}}ln\frac{\Lambda ^{2}}{m^{2}}\lesssim 1.\]
The suggested interpretation of the NLQFT then was (i) It is an effective
theory valid up to energy scale $\Lambda $ ;(ii) For energy scales
beyond $\Lambda $, one needs to replace the NLQFT by a more fundamental
theory (of constituents) having its own (larger) mass scale $\Lambda '$
and a coupling $g'$ and such constituent fields that the condition
$0<Z'<1$ is satisfied again within the energy domain $\Lambda <E\lesssim \Lambda '$.

An equivalent discussion can be given even for theories for which
such a condition is not available. It is based on the work \cite{j01},
in which we have attempted to give a mathematically rigorous understanding
of the renormalization program using the framework of the NLQFT. There,
we were lead to a very similar condition if the renormalization program
is to be understood in a rigorous way. We can then impose exactly
similar interpretation on the NLQFT.

\section{Causality Violation at One Loop }

\subsection{for $2$ Particle $\rightarrow $ $2$ Particle}

We shall now consider the contributions to $H_1$, of (\ref{eq:H_1}) a quantity quantifying the causality violation in the theory, for the above process. There are two contributions to this term: (a) One set of contributions comes from the diagrams with a shadow propagator (see fig. 1(a)) and analogous diagrams in $\emph{t-}$ and the $\emph{u}-$ channels. These contribute to $\int d^4x \int d^4yS_2(x,y)$ but not to the other term. (b) The second contribution comes from the renormalization counterterms present in $\int d^4x \int d^4yS_2(x,y)$ and $\int d^4x \int d^4yT[S_1(x)S_1(y)]$. The counterterm is (partly) determined by the requirement that the causality violation vanishes as $\Lambda \rightarrow \infty$.\\
(a) We have evaluated these diagrams
in the massless limit and for non-zero mass. 

\vspace{0.3cm}
\begin{center}\includegraphics[  width=4in,
height=3in] {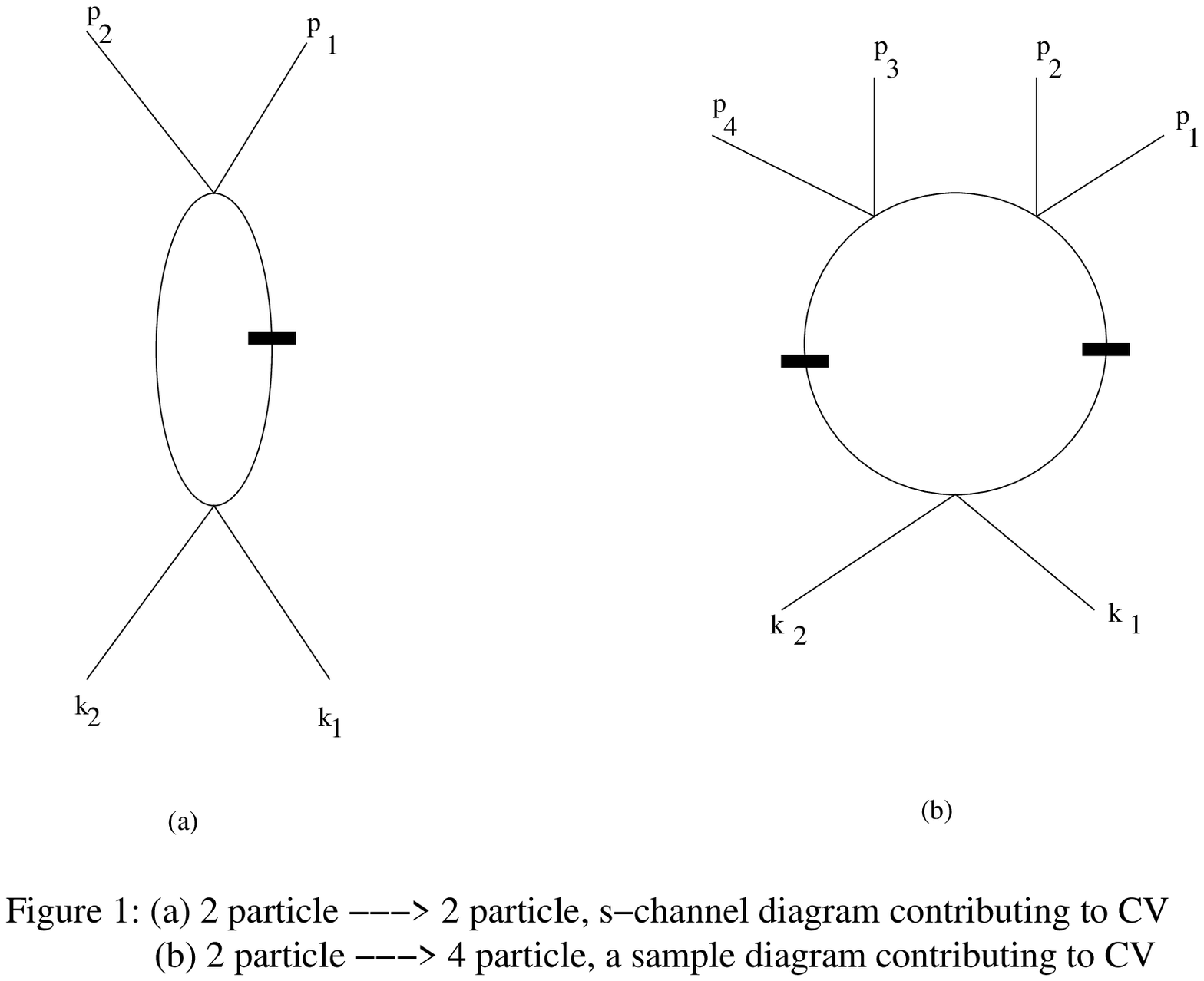}\end{center}
\vspace{0.3cm}

\subsubsection{Massless Limit}

After some involved calculations, we find that the amplitude for s-channel
diagram in the massless limit is given by\begin{equation}
\Gamma _{0}(s)= \frac{9\lambda ^2}{4\pi^2} \sum _{n=0}^{\infty }\frac{\left(\frac{s}{\Lambda ^{2}}\right)^{n}\left(1-\frac{1}{2^{n}}\right)}{n((n+1)!)}.\label{eq:schannel}  \end{equation}
The above expression, a convergent series, is particularly useful when $s \ll \Lambda ^2 $. In addition, there are the $t$-channel and the $u$-channel diagrams each of which is given by an expression (\ref{eq:schannel})  with $s \rightarrow t $ and $s \rightarrow u $ respectively. The sum of three such diagrams, for the case $s \ll \Lambda ^2 $ is given by,
\begin{eqnarray}  
\Gamma _{0}(s)+\Gamma _{0}(t)+\Gamma _{0}(u)& = &  \frac{9\lambda ^2}{4\pi^2}\left\{3 ln 2+ \frac{1}{4\Lambda^2}(s+t+u) +O\left(   \frac{s^2}{\Lambda^4 } \right)\right\}\\
& = & \frac{9\lambda ^2}{4\pi^2}\left\{3 ln 2+\frac{1}{\Lambda^2}(m^2) +O\left( \frac{s^2}{\Lambda^4 },\frac{t^2}{\Lambda^4 },\frac{u^2}{\Lambda^4 } \right)\right\}
\end{eqnarray}

The total causality violating amplitude is thus given by sum of the above result and the renormalization counterterm. One effect of the latter is to subtract out from \[\Gamma _{0}(s)+\Gamma _{0}(t)+\Gamma _{0}(u)\] the contribution that does not vanish away as $\Lambda \rightarrow \infty$. As explained in the section 2.2, we shall fix the counterterm further by assuming that CV vanishes at low momenta\footnote{Please recall the comment in section 2.2.}  ($s=t=u=0$). We thus obtain the net causality violating amplitude:
\begin{equation}
\frac{9\lambda ^{2}}{64{\pi}^2}\left(\frac{s^{2}+t^{2}+u^{2}}{\Lambda ^{4}}\right).\end{equation}
and thus is a \emph{quadratic} function of the basic momentum variables $s,t$ and $u$. Consequently, it grows much more slowly than expected with energy at low energies. One would therefore have to go to much higher energies to see causality violation than one would naively expect on the basis of dimensional arguments. In the section 4, we shall give a general argument to show that this must be so to all orders.

We shall now study the result for energies comparable to $\Lambda$. To see the behavior, we focus our attention on $\Gamma _0(s)$. We note that the series,
\begin{equation}
\Gamma _{0}(s)= \frac{9\lambda ^2}{4\pi^2} \sum _{n=0}^{\infty }\frac{\left(\frac{s}{\Lambda ^{2}}\right)^{n}\left(1-\frac{1}{2^{n}}\right)}{n((n+1)!)}.\end{equation}
for ${s}\sim {\Lambda ^2}$ grows exponentially. To see this, we introduce\\
$$A(\xi) \equiv  \sum_1^{\infty }\frac{\xi^n}{n(n+1)!} $$
We can then easily verify the following bounds for any $ \xi > 0$,\\
$$\left(\frac{e^{\xi}}{\xi^2}-\frac{1}{\xi^2}-\frac{1}{\xi}-1/2\right)< A(\xi)< \frac{1}{9}\left( \frac{e^{3\xi}}{\xi^2}-\frac{1}{\xi^2}-\frac{3}{\xi}-9/2\right) $$
[The lower bound is sufficient to conclude an exponential growth. The upper bound is a rather loose bound and can be made stricter, but that is not required].
We thus find that as $s \gtrsim \Lambda^2  $, there is an exponentially rising causality violation from the $s$-channel diagram.

\subsubsection{Non-zero Mass}

For completeness, we shall briefly discuss the case of $m \neq 0$; even though it does not lead to any substantial changes in conclusions. After invoking Schwinger parameterization and some calculations, we obtain the s-channel amplitude for this case in the following
integral form:\begin{equation}
\Gamma (s)\equiv \frac{9\lambda ^{2}}{4\pi ^{2}}\int _{0}^{1/2}dx\int _{\frac{1}{1-x}}^{\frac{1}{x}}\frac{d\tau }{\tau }exp\left\{ -\frac{\tau }{\Lambda ^{2}}\left(m^{2}-x(1-x)s\right)\right\} .\label{eq:Gammas} \end{equation}
 Expanding in powers of $s$ we get from the above expression \begin{equation}
\Gamma (s)=\sum _{n=0}^{\infty }\gamma _{n}(m,\Lambda )s^{n},\end{equation}
 where first three $\gamma _{n}$ are given by\begin{equation}
\gamma _{0}(m,\Lambda )=\frac{9\lambda ^{2}}{4\pi ^{2}}\int _{0}^{1/2}dx\int _{\frac{1}{1-x}}^{1/x}\frac{d\tau }{\tau }e^{-\frac{m^{2}\tau }{\Lambda ^{2}}},\end{equation}
 \begin{equation}
\gamma _{1}(m,\Lambda )=\frac{9\lambda ^{2}}{4\pi ^{2}}\int _{0}^{1/2}dx\int _{\frac{1}{1-x}}^{1/x}d\tau \frac{x(1-x)}{\Lambda ^{2}}e^{-\frac{m^{2}\tau }{\Lambda ^{2}}},\end{equation}
\begin{equation}
\gamma _{2}(m,\Lambda )=\frac{9\lambda ^{2}}{4\pi ^{2}}\int _{0}^{1/2}dx\int _{\frac{1}{1-x}}^{1/x}\tau d\tau \frac{x^{2}(1-x)^{2}}{2\Lambda ^{4}}e^{-\frac{m^{2}\tau }{\Lambda ^{2}}}.\end{equation}
The contributions to the total amplitude from the $n=0$
term is cancelled by the renormalization counterterm. Since $s+t+u=4m^{2},$ the contribution from the  $n=1$ terms add up to a momentum independent term. $\gamma _{2}(m,\Lambda )$
could be further simplified in terms of incomplete $\Gamma $-functions
but the expression is lengthy and complicated. To the lowest order
it has term $(const.)s^{2}/\Lambda ^{4}$ and at next leading order
there are terms like $m^{2}s^{2}/\Lambda ^{6}$ and $ln(m^{2}/\Lambda ^{2})m^{2}s^{2}/\Lambda ^{6}$. 

We make a number of observations, arising from the above discussion, from the point of view of their future relevance:
\begin{itemize}   \item  We note the existence of $m=0$ limit in the expression (\ref{eq:Gammas}).
\item We note that the expression (\ref{eq:Gammas}) can be expanded in powers of $s$ at $m=0$.
\item We can suspect the exponential-like growth of (\ref{eq:Gammas}) by an inspection of the integrand in (\ref{eq:Gammas}).
\end{itemize}

\subsection{Causality Violation at the One Loop level for the $2$ Particle $\rightarrow $
$4$ Particle process}

To obtain the appropriate quantification for CV for this process in the one-loop approximation, we simplify the causality condition (\ref{causal2}) using the  unitarity
condition (\ref{unitary2}). It reads,
\begin{eqnarray}
H_{2}(x,y,z) & = &    iS_{3}(x,y,z)-iS_{1}(x)S_{2}(y,z)-iS_{2}(x,y)S_{1}(z)-iS_{2}(x,z)S_{1}(y)\nonumber \\
& & -iS_1(x)S_1(y)S_1(z)-iS_1(x)S_1(z)S_1(y)=0\label{causal2'}
\end{eqnarray}

We find that for $2$ particle $\rightarrow $
$4$ particle process, the causality condition $H_2=0$ is violated by all the diagrams
with \emph{two} shadow propagators (see for example fig. 1(b)). To facilitate the calculation of CV in terms of the S-matrix elements, we use $H_{2}$ of (\ref{eq:H_2}). We find that $H_{2}$ consists of all diagrams for the 1-loop S-matrix with two barred lines. For concreteness, we shall first focus on the diagram of fig. 1(b). For this diagram
the limit $m\rightarrow 0$ exists and in this limit the amplitude is given by%
\footnote{After doing Schwinger parameterization and integrating over loop momentum%
}\begin{equation}
\frac{27\lambda ^{3}}{2\pi ^{2}\Lambda ^{2}}\int _{0}^{1}d\tau _{1}d\tau _{2}\int _{1}^{\infty }\frac{d\tau _{3}}{(\tau _{1}+\tau _{2}+\tau _{3})^{2}}exp\left\{ \frac{k^{2}\tau _{2}}{\Lambda ^{2}}+\frac{p^{2}\tau _{3}}{\Lambda ^{2}}-\frac{(\tau _{2}k+\tau _{3}p)^{2}}{(\tau _{1}+\tau _{2}+\tau _{3})\Lambda ^{2}}\right\} \label{eq:6point}\end{equation}
 where $k=k_{1}+k_{2}$ and $p=p_{1}+p_{2}$. Other diagrams of this
type can be now evaluated using the crossing symmetry. In this 6-point
function no renormalization is required to be carried out. Therefore
the leading order term, though a constant, will not be removed by renormalization. Leading order term  (for $\frac{p^2}{\Lambda ^2}$,$\frac{p.k}{\Lambda ^2}$,$\frac{k^2}{\Lambda ^2}$ $<< 1 $) is
\begin{equation}
\frac{27\lambda ^{3}}{2\pi ^{2}\Lambda ^{2}}\int _{0}^{1}d\tau _{1}d\tau _{2}\int _{1}^{\infty }\frac{d\tau _{3}}{(\tau _{1}+\tau _{2}+\tau _{3})^{2}}=\frac{27\lambda ^{3}}{2\pi ^{2}\Lambda ^{2}}(-4ln2+3ln3),\end{equation}
 and since it is independent of any momenta this term remains the
same for all other diagrams (total 15 in number). The \emph{total} amplitude for the process is finite as $\Lambda \rightarrow \infty  $ and behaves as $\mbox{(a quadratic function of momenta)}^{-1}$. Thus, the \emph{relative} CV behaves as,
$$\frac{ \mbox{a quadratic function of momenta}}{\Lambda ^2}$$
and thus is expected to grow with energy.

On the other hand, for some of the various Lorentz invariants $ \sim \Lambda ^2$ (and having an appropriate  sign), the truncation of the exponent as in (\ref{eq:6point}) is not a good approximation. For example, for $k^2=s\sim \Lambda^2, p^2\sim \Lambda^2\,\mbox{and}\,(p_3+p_4)^2<<\Lambda^2\mbox{and fixed}$, the expression (\ref{eq:6point}) shows an exponential-like growth with $s$.

\section{A general argument for $\phi \phi \rightarrow \phi \phi $}

In this section, we shall attempt a simple way to generalize to all orders the result worked out for the $\phi \phi \rightarrow \phi \phi $ process in the previous section. For this purpose, we shall employ the results stated in the Appendix A regarding the infrared properties and analytic properties of the relevant amplitudes. The amplitude for the process and the CV in it is a Lorentz-invariant,
dimensionless function only of the  two independent Lorentz-invariants
(say $s$ and $t$) and of the parameters $m$ and $\Lambda $. Let
us parametrize the causality violating combination of S-matrix elements 
by $f(s,t,m^{2},\Lambda ^{2})$. We shall first enumerate the properties expected of $f$:

\begin{itemize}
\item $f(s,t,m^{2},\Lambda ^{2})$ vanishes as $\Lambda $$\rightarrow \infty $,
i.e. in the local limit;
\item $f(s,t,m^{2},\Lambda ^{2})$ is dimensionless;
\item $f(s,t,m^{2},\Lambda ^{2})$ will have the crossing symmetry.
\item $f(s,t,m^{2},\Lambda ^{2})$ is finite at $m=0$. We have given an argument regarding this in the Appendix A. To put the conclusions in it qualitatively, there is a cancellation of diagrams in the net CV amplitude; so that in the surving diagrams, the infrared behavior is softer by the presence of (sufficiently many) barred propagators  which have a better infrared behaviour. 
(Please note the barred propagator in section 2.1 for $m=0$)\footnote{This and the following property can be addressed to \emph{diagrammatically} by evaluating an arbitrary 1PI diagram and reducing it to the form of a Schwinger parametric integral such as of the form (\ref{eq:6point}). The presence of a shadow propagator restricts the corresponding Schwinger parameter $\tau_i $ to $[0,1]$ whereas the infrared divergence comes from $\tau_i \rightarrow \infty $. The property below about the expandability in $s,t$ requires the study of the coefficients of $s$ and $t$ in the exponent in the integral. Also note the observations made at the end of the section 3.1.}.  
\item $f(s,t,m^{2},\Lambda ^{2})$ can be expanded in a series in $s,t$ through $O(s^2)$ at $m=0$. See Appendix A for the reasons.
\item $f(s,t,m^{2},\Lambda ^{2})$ is ambiguous upto momentum independent subtractions that vanish as $\Lambda \rightarrow \infty$.
\end{itemize}
In this case, the leading dependence will be a \emph{linear} function
of $\frac{s}{\Lambda ^{2}}$ and $\frac{t}{\Lambda ^{2}}$. The only
function that also has the crossing symmetry is \[
f(s,t,m^{2},\Lambda ^{2})=a\left\{ \frac{s}{\Lambda ^{2}}+\frac{t}{\Lambda ^{2}}+\frac{u}{\Lambda ^{2}}\right\} +b\frac{m^{2}}{\Lambda ^{2}}+O\left(\frac{1}{\Lambda ^{4}}\right)\]
In view of $s+t+u=4m^{2}$, we find that \[
f(s,t,m^{2},\Lambda ^{2})=(b+4a)\frac{m^{2}}{\Lambda ^{2}}+O\left(\frac{1}{\Lambda ^{4}}\right)\]
In view of the last observation regarding an ambiguity in the renormalization counterterms,
we have,\[
f(s,t,m^{2},\Lambda ^{2})=O\left(\frac{1}{\Lambda ^{4}}\right)\]
modulo momentum-independent terms $O\left(\frac{1}{\Lambda ^{2}}\right).$Thus, the leading momentum-dependent CV in this process is of $O[\Lambda ^{-4}]$ i.e.
two powers of $\Lambda $ suppressed as compared to what one would expect.

\section{A possible scenario for escape from large scale causality violation at high energy scales}

The scale $\Lambda$ of the non-local standard model is, in principle, available  phenomenologically from various precison experiments such as $(g-2)$ of the muon \cite{js}, the precision tests of standard model \cite{aj} etc, i.e. from experiments conducted at energies $\ll \Lambda$ . As we had seen in sections 3 and 4, the causality violation can begin to
grow very rapidly with the energy scale as the latter becomes a substantial
fraction of the mass-scale $\Lambda $ of the theory, so determined, and grows exponentially beyond it.  It is possible
that such a causality violation is actually present at high enough
energies; in which case this will vindicate the use of non-local theory as the correct formulation of the physical theory, since
such a causality violation cannot be explained by a LQFT. On the other
hand, however, it is also possible that experiments detect no significant
violation of causality even at such energies. One need not, however, necessarily conclude from it that the use of the non-local field theory to represent fundamental physics need be abandoned. In this section,
we shall present a possible scenario that can provide an escape from
such a predicament. In fact, we shall argue that such a predicament is indicative of a presence of a finer structure underlying the present theory.

We shall find it useful to employ the scenario presented in \cite{j01_2} (which we have briefly reviewed in section 2.4) for the present
purpose also. In that work, we had argued that the condition $0<Z<1$
that is supposed to be satisfied by the wave-function renormalization
should in fact be taken literally in perturbation theory and can be
implemented within the context of the non-local field theory. It can be
used to predict the scale $\Lambda $ that is present in the non-local
formulation by a condition of the type\[
\frac{g^{2}}{16\pi ^{2}}ln\frac{\Lambda ^{2}}{m^{2}}\lesssim 1.\]
[A similar conclusion was drawn in \cite{j01} from a discussion regarding renormaliztion program in a renormalizable field theory (including the standard model) where a condition $0<Z<1$ may not be derivable]. It was further proposed that the non-local field theory is an effective field theory and the effective field theory should be
abandoned beyond energy scale $\gtrsim \Lambda $ where it must be
replaced by a more fundamental theory with another coupling $g'$
and mass scale $\Lambda '$. Such a scenario can also protect
a non-local theory from a "causality-catastrophe". 

We propose that should a large causality violation be unobserved at energies $\sim \Lambda$ (obtained from the above condition), it can be understood by invoking the above picture. In this picture, the calculations done in the non-local field theory at energy scales  $\sim \Lambda$ are no longer very accurate, as the substructure of the theory becomes important. At these energy scales, the non-local field theory, which is being looked upon as an \emph{effective} field theory, should be replaced by another non-local theory of its fundamental constituents having a new scale $\Lambda'>> \Lambda$ and a new coupling constant. It is viable that in this new setting the causality violation may turn out to be in fact small as the energies are $<<\Lambda'$.

In other words, in this view, the \emph{calculated} large causality violation in NLQFT at large energy scales is argued as an artifact of the approximation that replaces the underlying theory of fundamental constituents  by an effective non-local field theory.

The experimental observations regarding the CV at high enough energies may, in fact, enable one to distinguish between these two views regarding a non-local theory: (i) A \emph{fundamental} theory with a mass parameter $\Lambda$, (ii) an effective field theory valid upto a scale $\Lambda$.

\section{Appendix A}

In this appendix, we shall briefly deal with the infrared properties
of the causality violating amplitudes constructed out of $\textrm{H}_{\textrm{n}}(y,x_{1},......,x_{n})$
that we have discussed in section 2.2 (see Eq.(4)). We have employed
these properties in the section 4. On physical grounds, we can expect
that quantities that characterize causality violation such as these
should not have a sensitive dependence on $m$ as $m\rightarrow 0$,
that is, an absence of a mass singularity. We have already seen examples
of lack of mass singularities in the calculations we have done. For
example, in section 3.1, we have calculated the one loop amplitude
arising from $H_{1}$ and found it to have

\begin{itemize}
\item Contribution from diagrams necessarily having one barred line (which
have softer infrared properties);
\item No singularity at $m=0$ ( see section 3.1.1);
\item The amplitude is \emph{analytic} (in $s,t,u$) at $m=0$.
\item There is no imaginary part, as there are no physical (i.e. having
at least two smeared lines) intermediate states.
\end{itemize}
Further in section 3.2, we have seen the evaluation of the 1-loop
amplitude based on $H_{2}$. We found there that 

\begin{itemize}
\item the contribution came from graphs with \emph{at least two} barred
lines;
\item there is no singularity at $m=0$;
\item The amplitude is \emph{analytic} in the quadratic Lorentz invariants
at $m=0$. There is no imaginary part, as there are no physical intermediate
states.
\end{itemize}
We draw attention to the fact that the $H_{n}$'s contain \emph{specific}
combinations of several terms involving various $S_{m}$'s . In $H_{n}$'s,
there is a cancellation of diagrams: In the 4-point function of $H_{1}$,
the diagrams with only smeared propagator have cancelled out, while
in the 6-point function of $H_{2}$, diagrams with only smeared lines
and those with \emph{one} barred line have cancelled out. We shall
argue that while each term in $H_{n}$ may have infrared divergences/mass
singularities as $m\rightarrow 0$, the combinations above cannot.

To see the argument, we consider the following quantity,\[
S[g(x),m]S^{\dagger }[g(x),m]\equiv 1\, \, \, \, \, \, for\, \, all\, \, m\, \, including\, \, m\rightarrow 0\]
We note that the left hand side has a smooth limit as $m\rightarrow 0$
and consequently has no mass singularities in its matrix elements.
We now imagine a small \emph{local} variation \emph{}in $g(x)$:$g(x)\rightarrow g(x)+\delta g(x)$
in the first factor. We argue that \emph{local} variations such as
these cannot affect the infrared properties of the theory, which necessarily
arise from large distances. Thus, the following operator: $S[g(x)+\delta g(x),m]S^{\dagger }[g(x),m]$
or equivalently the operator\[
O(y)\equiv i\frac{\delta S[g,m]}{\delta g(y)}S^{\dagger }[g,m]\]
has a smooth limit as $m\rightarrow 0$, i.e. has no mass singularity
in its matrix elements. We note the $H_{n}(y,x_{1},.....x_{n})$ has
been defined as,\[
H_{n}(y,x_{1},.....x_{n})\equiv \left.\frac{\delta ^{n}}{\delta g(x_{1})\delta g(x_{2}).....\delta g(x_{n})}O(y)\right|_{_{g=0}}\]
and the causality violating combinations have been constructed out
of these.

Next, we shall briefly deal with the analytic nature of the diagrams.
It turns out that we can get at some of the analytic properties \emph{without}
detailed analysis of diagrams. Consider\[
S[g,m]S^{\dagger }[g,m]=1\]
We differentiate with respect to $g(y)$ to obtain,

\[
\frac{\delta S[g,m]}{\delta g(y)}S^{\dagger }[g,m]+S[g,m]\frac{\delta S^{\dagger }[g,m]}{\delta g(y)}=0\]
Thus, $O(y)\equiv i\frac{\delta S[g,m]}{\delta g(y)}S^{\dagger }[g,m]$
is a hermitian operator \cite{bs1}. Now consider the diagonal matrix elements
of $O(x)$ between two 2-scalar states $|\alpha >\equiv |k_{1},k_{2}>$.
Then, \[
<\alpha |O(y)|\alpha >=<\alpha |i\frac{\delta S[g,m]}{\delta g(y)}S^{\dagger }[g,m]|\alpha >=a\, real\, quantity\]

Now, matrix elements of $H_{n}$ (analogues of (9) and (10)) are constructed
out of \[
\frac{\delta }{\delta g(x_{1})}\frac{\delta }{\delta g(x_{2})}.....\frac{\delta }{\delta g(x_{n})}<\alpha |i\frac{\delta S[g,m]}{\delta g(y)}S^{\dagger }[g,m]|\alpha >=a\, real\, quantity\]
by real operations (see e.g. Eq.(10)). Thus, the quantity $<\alpha |H_{n}|\alpha >$
does not develop imaginary part and the associated singularities due
to contributions from physical intermediate states and thus cannot
have singularities. This implies that the amplitudes so calculated
$f(s,t,m,\Lambda )$ are free of singularities at $t=0$ for any $m\geq 0$.
This, in particular, rules out the terms of the type $\frac{slns}{\Lambda ^{2}}$
at $m=0$. Crossing symmetry rules out terms such as $\frac{tlnt}{\Lambda ^{2}}$.

\end{document}